\documentclass[showpacs,preprintnumbers,amsmath,amssymb]{revtex4}
\usepackage{nicefrac}
\usepackage{amssymb}
\usepackage{amsmath}
\usepackage{mathrsfs}
\usepackage{graphics}
\usepackage{epsfig}
\usepackage{graphicx}
\usepackage{subfigure}
\usepackage{bm}
\usepackage{nicefrac}

\def\beq{\begin{equation}}
\def\eeq{\end{equation}}
\def\baq{\begin{eqnarray}}
\def\eaq{\end{eqnarray}}
\def\a{\alpha}

\def\be{\begin{equation}}
\def\ee{\end{equation}}
\def\bea{\begin{eqnarray}}
\def\eea{\end{eqnarray}}

\def\hnl{h_{\rm NL}}
\def\fnl{f_{\rm NL}}

\def\nfnl{n_{f_{\rm NL}}}
\def\gnl{g_{\rm NL}}

\def\ngnl{n_{g_{\rm NL}}}
\def\taunl{\tau_{\rm NL}}

\def\hnl{h_{\rm NL}}

\def\ntaunl{n_{\tau_{\rm NL}}}
\def\vp{\varphi}

\begin{document}

\unitlength = 1mm
\title{ Vector instabilities and self-acceleration in the decoupling limit of massive gravity}
\author{Gianmassimo Tasinato$^{(1)}$, Kazuya Koyama$^{(1)}$, Gustavo Niz$^{(1,2)}$}
\affiliation{%
~ \\
 $^{(1)}$ Institute of Cosmology $\&$ Gravitation, University of Portsmouth,
\\\hskip0.2cm 
 Portsmouth, PO1 3FX, United Kingdom\\
$^{(2)}$ Departamento de F\'{\i}sica, Universidad de Guanajuato,\\
 DCI, Campus Le\' on, C.P. 37150, Le\' on, Guanajuato, M\' exico.
}%

\begin{abstract}
\noindent
We investigate 
vector contributions to the Lagrangian of $\Lambda_3-$massive gravity in the decoupling limit, the less explored sector of this theory. The main purpose is to understand the stability of maximally symmetric 
vacuum solutions. Around self-accelerating configurations,
vector degrees of freedom become strongly coupled since their kinetic terms vanish, so their dynamics is controlled by higher order interactions. Even in the decoupling limit, the vector Lagrangian contains an infinite number of terms. We develop a systematic method to covariantly determine the vector Lagrangian at each order in perturbations, fully manifesting the symmetries of the system. We show that, around self-accelerating solutions, the structure of higher order $p$-form Galileons arise, avoiding the emergence of a sixth BD ghost mode. However, a careful analysis shows that there are directions along which the Hamiltonian is unbounded from below. This instability can be interpreted as one of the available fifth physical modes behaving as a ghost. Therefore, we conclude that self-accelerating configurations, in the decoupling limit of $\Lambda_3$-massive gravity, are generically unstable.

\end{abstract}
\maketitle

\smallskip





\def\tr{{\rm tr}}
\def\hnl{h_{\rm NL}}
\def\fnl{f_{\rm NL}}
\def\nfnl{n_{f_{\rm NL}}}
\def\gnl{g_{\rm NL}}
\def\ngnl{n_{g_{\rm NL}}}
\def\taunl{\tau_{\rm NL}}
\def\hnl{h_{\rm NL}}
\def\ntaunl{n_{\tau_{\rm NL}}}
\def\vp{\varphi}
\def\ck{{\cal K}}

\newcommand{\inl}{i_{\rm NL}}
\newcommand{\half}{\frac{1}{2}}
\newcommand{\bright}{\begin{flushright}}
\newcommand{\eright}{\end{flushright}}
\newcommand{\bminip}{\begin{minipage}}
\newcommand{\eminip}{\end{minipage}}
\newcommand{\bcent}{\begin{center}}
\newcommand{\ecent}{\end{center}}
\renewcommand{\thefootnote}{\fnsymbol{footnote}}
\newcommand{\nnb}{\nonumber}
\newcommand{\reflef}{(\ref}
\newcommand{\MP}{M_{\rm P}}
\newcommand{\BBbox}{\mbox{\Large $\sqcap$}\hspace{-1.0em}\mbox{\Large $\sqcup$}}\newcommand{\lmd}{\lambda}
\newcommand{\Lmd}{\Lambda}
\newcommand{\gsim}{\mbox{\raisebox{-.3em}{$\;\stackrel{>}{\sim}\;$}}}
\newcommand{\lsim}{\mbox{\raisebox{-.3em}{$\;\stackrel{<}{\sim}\;$}}}





\section{Introduction}

One of the deepest problems in  theoretical cosmology is to provide a satisfactory  explanation for the  observed acceleration of our universe.
It is intriguing to ask whether this phenomenon may be due to a non-standard dynamics of gravity at large distances. Einstein's General Relativity cannot lead to cosmological acceleration without the addition of an energy momentum tensor of the suitable form.
The simplest choice is a positive cosmological constant term, whose small value is, however,   unnatural from an effective quantum field theory perspective. Modifications
to General Relativity might achieve acceleration by exploiting self-interactions of gravitational degrees of freedom, without adding an energy momentum tensor. This phenomenon is dubbed cosmological self-acceleration.

An early explicit realization of this idea is the model of Dvali, Gabadadze, and
Porrati (DGP) \cite{Dvali:2000hr}. In this set-up, a brane is embedded  in a higher dimensional space-time with extra-dimensions of infinite volume. Four dimensional gravity is recovered at small scales thanks to an Einstein-Hilbert term localised on the brane. Above a cross-over scale, which depends on a combination of the four and higher dimensional Planck masses, gravity becomes higher dimensional, and this leads to cosmological self-acceleration. This scenario is appealing for its simplicity, and for its clear geometrical interpretation. However, the dynamics of cosmological fluctuations    around self-accelerating solutions is plagued by ghost instabilities; some of the dynamical modes have kinetic terms with the wrong sign \cite{Koyama:2005tx}. A way out to this conclusion can be achieved by modifying and generalizing this set-up to develop scenarios with the right features to avoid the emergence of a ghost. A possibility is to adopt the Galileon proposal \cite{
Nicolis:2008in}. An interesting model, so-called $\Lambda_3$
massive gravity \cite{deRham:2010ik}, has been developed based on this suggestion,
 and admits self-accelerating configurations, with the scale of acceleration set by the graviton mass.  This is a generalization of Fierz-Pauli massive gravity with the addition of higher order graviton self-interactions that are free of the Boulware-Deser 'sixth' ghost mode, which plagues the original Fierz-Pauli theory
\cite{ghost}. Around a flat Minkowski background, the theory propagates five healthy degrees of freedom: two helicity 2, two helicity 1, and a scalar mode. However, one (or more) of these five modes may become a ghost around solutions that are different from Minkowski space.

In the so-called decoupling limit \cite{ArkaniHamed:2002sp}, the structure of self-accelerating configurations is relatively easy to study, and the theory exhibits the explicit Galileon symmetry in the scalar sector, resulting in an attractive model from the point of view of stability under classical and quantum corrections (see \cite{Hinterbichler:2011tt,deRham:2012az} for reviews). If instabilities are found in the regime well
described by the decoupling limit, the theory is certainly sick.
On self-accelerating cosmological solutions, it has been shown that vector degrees of freedom become strongly coupled, being characterized by vanishing kinetic terms
\cite{deRham:2010tw, Koyama:2011wx}. Our purpose is to analyze in detail the Lagrangian controlling these
vector modes, the less explored sector of the theory so far, and to understand whether they can lead to instabilities around cosmological configurations. Even in the decoupling limit, the Lagrangian for  vector degrees of freedom contains an infinite number of terms. Extending the approach introduced in \cite{Koyama:2011wx}, we are able to develop a systematic  method that
allows to determine in a  covariant way the vector Lagrangian at each order in perturbations,  fully manifesting the symmetries
of the system.
 In order to study the  role of vectors  in the dynamics, one can proceed in different ways. In \cite{Koyama:2011wx} we showed that the inclusion of a background vector, while maintaining the qualitative properties of the background solutions, removes the strong coupling of vector perturbations, but forces one of the five degrees of freedom to become a ghost thus ruling out self-accelerating solutions
 with a background vector switched on.
  More recently,
   a qualitatively similar result was obtained away from the decoupling limit, by turning on background anisotropy and considering
   Bianchi-type solutions \cite{DeFelice:2012mx}.   
   In this work, instead, we focus on solutions with no background vector field and analyze the dynamics of perturbations. Scalar-vector interactions, appearing in the Lagrangian at order higher than two in perturbations, are the key points for our analysis.
 At third order in perturbations, these couplings
form the so-called $p$-form combination \cite{Deffayet:2010zh} (see also \cite{Zhou:2011ix}),
 and  lead to equations of motion with at most two time derivatives.
  The same structure continues at fourth order in perturbations, after  a suitable field redefinition has been performed.
  Besides showing this explicitly at fourth order, we provide strong arguments indicating  that the same pattern continues at higher orders in perturbations; this confirms from a new perspective  that the theory does not contain a sixth ghost BD mode around
  self-accelerating solutions.
  On the other hand, a Hamiltonian analysis of the resulting system including higher order perturbations reveals that the Hamiltonian is unbounded from below, signaling an instability which in turn can be understood as
    a ghost within the available five propagating modes.
 Therefore, our analysis shows, by a direct study of the Lagrangian structure at higher order in perturbations, that self-accelerating solutions in   $\Lambda_3$ massive gravity are indeed unstable.

Our analysis of the vector Lagrangian in the decoupling limit is sufficiently simple that allows to physically understand  the origin
of the instabilities around self-accelerating configurations. At the same time makes manifest that the vector Lagrangian in this theory
organizes in a very non-trivial way at higher order in perturbations, avoiding the appearance of a ghostly Boulware-Deser 'sixth mode' by
forming combinations corresponding to $p$-form Galileon combinations. This allows to appreciate from a new perspective how a theory (in this particular case $\Lambda_3$ massive gravity) can subtly avoid Ostrogradski-type instabilities \cite{ostro,Woodard:2006nt},
though it leads to a different type of instabilities around self-accelerating backgrounds due to the lack of the vector kinetic term.


 \smallskip
This work is organized as follows.  After starting with a brief review of $\Lambda_3$ massive gravity in decoupling limit in section \ref{sec-prel}, we consider in
section \ref{sec-msna} the theory in the minimal set-up, $\alpha_3=\alpha_4=0$ (see the next section \ref{sec-prel} for a definition of these parameters), in the presence of a bare cosmological constant.  We are able to classify all the maximally symmetric solutions of this theory, as a function of the cosmological constant. We show that no healthy maximally symmetric solutions de Sitter or anti-de Sitter solutions exist in this case. In section \ref{sec-mswa} we generalize our discussion to the full theory with arbitrary $\alpha_3$ and $\alpha_4$, reaching the same conclusions. Section \ref{sec-disc} is devoted to a summary of our results,  followed by three technical appendixes.

\section{Preliminaries: $\Lambda_3$ massive gravity in a decoupling limit}\label{sec-prel}


The $\Lambda_3$ theory of massive gravity \cite{deRham:2010ik} is designed to remove the Boulware-Deser ghost by an uplift of a decoupling-limit construction of massive gravity, which does not propagates a sixth scalar degree of freedom.
The resulting Lagrangian is then the following
\be
{\cal L}\,=\,\frac{M_P^2}{2}\,\sqrt{-g}\,\left(R- {\cal U}(g)  -12 \Lambda_{cc}\right),
\ee
where $R$ is the Ricci scalar, and ${\cal U}(g)$ is a potential term for the graviton that will be explicitly shown later. We add a pure cosmological constant $\Lambda_{cc}$ to the original formulation \cite{deRham:2010ik}, but avoid any additional energy momentum tensor, since we focus on  maximally symmetric vacuum configurations. These solutions are the simplest examples of cosmological solutions in the theory which, nevertheless, exhibit a rich behavior with respect to cosmological fluctuations. Cosmological solutions in this theory, and in the related  ghost-free bigravity models, have been extensively studied \cite{deRham:2010tw,Koyama:2011yg,Koyama:2011xz,Gumrukcuoglu:2011ew,D'Amico:2012pi,Gumrukcuoglu:2011zh,cosmorefs}. In order to write down the potential ${\cal U}(g)$, it useful  to express the graviton in a diffeomorphism invariant way, by means of the St\"uckelberg trick; hence writing the metric as
\be\label{stutrick}
g_{\mu\nu}\,=\,\eta_{\mu\nu}+\frac{h_{\mu\nu}}{M_P}\,=\,H_{\mu\nu}+\eta_{\alpha \beta}\,\partial_\mu \phi^\alpha \partial_\nu \phi^\beta,
\ee
where the four St\"uckelberg fields $\phi^\alpha$ transform as scalars under Lorentz transformations of the physical metric $g_{\mu\nu}$, while transforming as four vectors from the fiducial metric $\eta_{\alpha \beta}$ perspective. Greek indeces starting from $\mu$ indicate quantities evaluated in physical space, while greek indeces starting from $\alpha$ indicate quantities evaluated in the fiducial space. Notice that we expand the metric $g_{\mu\nu}$ about a flat background $\eta_{\mu\nu}$ in physical space. Using Helmoltz theorem in the fiducial space, the helicity one, $A_\beta$, and the helicity zero, $\pi$, modes of the St\"uckelberg fields $\phi^\alpha$ can be extracted by the expansion
\be
\phi^\alpha\,=\, x^\alpha-\frac{m\,\eta^{\alpha \beta}\,A_\beta }{\Lambda_3} -\frac{\eta^{\alpha \beta}\,\partial_\beta \pi}{\Lambda_3}\,,
\label{phidec1}\ee
where
\be
\Lambda_3\,=\,\left( m^2 M_{Pl}\right)^{\nicefrac{1}{3}},
\ee
corresponds to the cut-off scale for this theory. The powers of $m$ and $\Lambda_3$ in eq.~(\ref{phidec1}) are included in such a way to provide, after a suitable diagonalization, canonical kinetic terms for the tensor, vector and scalar degrees of freedom around Minkowski space-time. In the remaining part of this section, we set the vector fluctuations to zero, but will return to study them in detail in the following sections.

We are interested in studying the so-called decoupling limit of $\Lambda_3$ massive gravity, which is defined as
\be\label{defdec1}
m\to 0\,,\hskip1cm M_{Pl}\to \infty\,,\hskip1cm \Lambda_3\,=\,{\text{fixed}}\,.
\ee
Moreover, in this limit, we demand that the cosmological constant $\Lambda_{cc}$ survives such that
\be M_{P}^2 \Lambda_{cc}\to \Lambda_3^2 \lambda_0\,\;\;, \hskip1cm \lambda_0\,=\,{\text{fixed }}\,,\ee
with $\lambda_0$ of dimensions of mass to the fourth.
The resulting theory describes well the physics up to energies of  order $\Lambda_3$, where new higher dimensional operators arise (suppressed by new scales larger than $\Lambda_3$). In this work we focus on scales smaller or equal to $\Lambda_3$.
When convenient, we use the abbreviations $$\Pi_{\mu\nu}=\partial_\mu \partial_\nu \pi \hskip1cm {\text{and}}\hskip1cm\Pi=\partial_\mu \partial^\mu \pi\,\,.$$

In the decoupling limit, it is equivalent to raise and lower indices with the physical or the fiducial metric; both reduce to flat space since the effect of $h_{\mu\nu}$ can be neglected in the limit of infinite Planck mass.
In order to render the formulae less cumbersome, we set the scale $\Lambda_3=1$;
if desired, it is straightforward to re-instate it by means of dimensional arguments.
Then the tensor $H_{\mu\nu}$ appearing in eq. (\ref{stutrick}) reads (neglecting for the time being the effect of the vectors)
\be
H_{\mu\nu}\,=\,\frac{h_{\mu\nu}}{M_P}+2  \Pi_{\mu\nu}-\eta^{\alpha \beta}\Pi_{\mu \alpha} \Pi_{\nu \beta}.
\ee
It is convenient to define a quantity
\be
\ck_{\mu}^{\nu}\,=\,\delta_\mu^\nu-\sqrt{\delta_\mu^\nu-H_\mu^\nu},
\ee
which has the property that ${\ck_{\mu\nu}}|_{_{h_{\mu\nu,\,A_\lambda}=0}}\,=\,\Pi_{\mu\nu}$.
By means of this quantity, we consider the following potential \cite{deRham:2010ik}
\be
{\cal U}= -m^2\left[{\cal U}_2+\alpha_3\, {\cal U}_3+\alpha_4\, {\cal U}_4\right],
\label{potentialU}
\ee
with
\def\mK{{\mathcal K}}
\bea
{\cal U}_2&=&(\tr\mK)^2-\tr (\mK^2),\nonumber \\
{\cal U}_3&=&(\tr \mK)^3 - 3 (\tr \mK)(\tr \mK^2) + 2 \tr \mK^3,\nonumber \\
{\cal U}_4 &=& (\tr \mK)^4 - 6 (\tr \mK)^2 (\tr \mK^2)
+ 8 (\tr \mK)(\tr \mK^3) + 3 (\tr \mK^2)^2 - 6 \tr \mK^4 \,,\nonumber
\eea
where $m$ has dimension of a mass, while $\alpha_3$ and $\alpha_4$ are dimensionless parameters. By construction, this potential leads to the following ghost free Lagrangian for tensor and scalar modes in a decoupling limit  \cite{deRham:2010ik}
\bea\label{lagsim1}
{\cal L}_{h_{\mu\nu},\;\pi}&=&-\frac12 \,h^{\mu \nu} {\cal E}_{\mu \nu}^{\alpha \beta} h_{\alpha\beta}
- 3 \,\lambda_0\, h +h^{\mu\nu} \left( {X}^{(1)}_{\mu\nu} +{X}^{(2)}_{\mu\nu} +{X}^{(3)}_{\mu\nu} \right),
\eea
where ${\cal E}_{\mu \nu}^{\alpha \beta}$ is the operator acting on $Z_{\alpha \beta}$ as
\be
 {\cal E}_{\mu \nu}^{\alpha \beta} Z_{\alpha\beta}\,=\,-\frac12\,\left(
 \Box Z_{\mu\nu}-\partial_\mu \partial_\alpha Z^\alpha_\nu-\partial_\nu \partial_\alpha Z^\alpha_\mu
 +\partial_\mu \partial_\nu Z^\alpha_\alpha-\eta_{\mu\nu} \Box Z^\beta_\beta +\eta_{\mu\nu} \partial_\alpha
 \partial_\beta Z^{\beta \alpha}
 \right).
\ee
The expressions for the $X^{(i)}_{\mu\nu}$ are given by
\bea
X^{(1)}_{\mu\nu}&=&
 \Big[ \eta_{\mu\nu} \Pi
 -\Pi_{\mu\nu}
 \Big]\,,\nonumber
\\
X^{(2)}_{\mu\nu}&=& \left(1+3 \alpha_3\right)\Big[
\Pi_\mu^\lambda \Pi_{\lambda \nu}-\Pi \Pi_{\mu\nu}+\frac12\eta_{\mu\nu}\left( \Pi^2-\Pi_{\mu}^\nu \Pi_\nu^\mu\right)
\Big]\label{xiexp}\,,
\\
X^{(3)}_{\mu\nu}&=&-\frac12\,
 \left(\alpha_3+3\alpha_4\right)\Big[
 6 \Pi_{\mu \nu}^3- 6 \Pi\,\Pi_{\mu\nu}^2+3 \Pi_{\mu\nu} \left( \Pi^2-\Pi_\mu^\nu\,\Pi_\nu^\mu \right)
 -\eta_{\mu\nu}
 \left(
 \Pi^3-3 \Pi \Pi_\mu^\nu\,\Pi_\nu^\mu+2 \Pi_\mu^\nu\,\Pi_\nu^\rho \,\Pi_\rho^\mu
 \right)
  \Big]\nonumber\,.
\eea

This Lagrangian is at most quadratic in $h_{\mu\nu}$ and the scalar $\pi$ couples to $h_{\mu \nu}$ through higher derivative interactions, which lead to equations of motion with at most two time derivatives. Notice that the Lagrangian (\ref{lagsim1}) contains a linear term proportional to $\lambda_0\, h$, which indicates that Minkowski space, with $\phi^\alpha=x^a$, is not a solution of the background field equations. In order to remove this linear term, one can set the cosmological constant to zero ($\lambda_0=0$), or consider more general background configurations, as we will do in the next section.

\section{Maximally symmetric configurations with  $\alpha_3=\alpha_4=0$}\label{sec-msna}

In this section, we consider a generalization of the maximally symmetric configurations found in \cite{deRham:2010tw} including a cosmological constant, and discuss the dynamics of fluctuations around them.  We focus here on the case $\alpha_3=\alpha_4=0$, postponing the analysis of the general case to section \ref{sec-mswa}. The novelty of our discussion is the detailed study of the vector fluctuations around cosmological solutions. These cosmological backgrounds are solutions in the decoupling limit, so one might wonder whether there exist maximally symmetric configurations of the full theory which reduce to the ones we study here. The answer is affirmative; we explicitly showed this fact in \cite{Koyama:2011wx} for the solutions found in \cite{Koyama:2011yg,Koyama:2011xz}, and by the same methods, it should be straightforward to investigate other configurations found in the literature as \cite{Gumrukcuoglu:2011ew} (see also \cite{D'Amico:2012pi}).

\subsection{Properties of tensor and scalar fluctuations}
An important property of the decoupling limit (\ref{defdec1}) is that it does not require the scalar $\pi$ to be small.   The Lagrangian given in (\ref{lagsim1}) is correct no matter what size the scalar $\pi$ is. In other words, the decoupling limit Lagrangian is not a perturbative expansion in field fluctuations, but a perturbative expansion in the small scales $m$ and $1/M_{Pl}$, while keeping $\Lambda_3$ fixed. This is different from standard perturbation theory, in which instead the quantities involved are small and only contributions up to a given order in perturbations are included. We can exploit this fact to straightforwardly determine maximally symmetric solutions in this set-up. Let us parameterize the $h_{\mu\nu}$ and $\pi$ degrees of freedom as
\bea\label{shiftquan}
h_{\mu\nu}&=&  -\frac{H^2}{2}\, x^2\,\eta_{\mu\nu}+\hat{h}_{\mu\nu}\,,\\
\pi&=&\frac{\left(1-c_0\right)}{2}\,x^2+\hat{\pi}\label{pishift1}\,,
\eea
where $x^2\,=\,x^{\mu} x^{\nu} \eta_{\mu\nu}$, and $c_0$ is a constant parameter. $H^2$ is a quantity with dimension of mass to the cube (this is the quantity dubbed $M_{Pl} H^2$ in \cite{deRham:2010tw}, with $H^2$ of dimension mass squared; in their notation, the scale of $H^2$ is set by the graviton mass $m^2$). A non-trivial profile for the vector could be included, but we have already studied the subject in detail in \cite{Koyama:2011wx} (at least for  spherically symmetric set-ups).
By a suitable choice of $c_0$ and $H$, the previous expressions solve the equations of motion for
 the lagrangian (\ref{lagsim1}), so that the background metric and Stuckelberg fields would then be given by
\bea
g_{\mu\nu}^{(0)}&=&\left( 1- \frac{H^2}{2\,M_{Pl}} x^2\right) \eta_{\mu\nu} \label{metrdec1}\,,\\
\phi_{(0)}^\alpha&=&c_0 \,x^\alpha \label{stuckdec1}\,.
\eea
and $\hat{h}_{\mu\nu}$ and $\hat \pi$ will be the quantities playing the role of our helicity two and helicity zero 'small' fluctuations around this background.

The metric corresponds to a maximally symmetric space: (A)dS for (negative) positive $H^2$. In the decoupling limit, the metric of a maximally symmetric space with vanishing spatial curvature can be reduced to the form (\ref{metrdec1}) by appropriate change of coordinates;
a background profile for the St\"uckelberg fields (\ref{stuckdec1}) allows us to solve the corresponding equations of motion as we will to discuss in what follows.

\smallskip
Plugging the parameterization (\ref{shiftquan}) into the Lagrangian for tensor and scalar
fluctuations, eq.~(\ref{lagsim1}) with $\alpha_3=\alpha_4=0$, and using the following
useful relations, which are valid up to total derivatives when included in the action
\bea
{\cal E}_{\mu\nu}^{\;\;\rho\sigma}  \left(
{\eta}_{\rho\sigma}\,x^2\right)&=&
6 \,\eta_{\mu\nu}\,,
\\
x^2\,\eta^{\mu\nu}\,\left[ \eta_{\mu\nu} \,\Box \hat \pi-\partial_\mu \partial_\nu \,\hat \pi\right]&=& 24 \hat \pi\,,
\\
x^2 \left[ (\partial_\mu \partial_\nu \pi)  (\partial^\mu \partial^\nu \hat \pi)  -( \Box \hat \pi )^2 \right]&=&-6 \hat\pi \Box \hat\pi\,,
\label{urel3} \eea
one finds
\bea\label{lagdec2}
{\cal L}_{dec}&=& -\frac12 \,\hat{h}^{\mu \nu} {\cal E}_{\mu \nu}^{\alpha \beta} \hat{h}_{\alpha\beta}
- 3 \left[  \lambda_0 -   H^2-\left(1-c_0\right) \left(2-c_0\right)\right]\,  \hat{h}\nonumber
\\
&&
-12   H^2 (3-2 c_0) \hat{ \pi} +\hat{h}^{\mu\nu} {X}_{\mu\nu}-6  H^2\,\hat{\pi} \Box \hat{\pi}\,.
\eea
The expression for $\hat{X}_{\mu\nu}$ is given by
\bea
\hat{X}_{\mu\nu}&=&
 \Big[ (3-2 c_0)\eta_{\mu\nu} \hat \Pi
 -(3-2 c_0) \hat \Pi_{\mu\nu}-\hat \Pi \hat \Pi_{\mu\nu}+\hat \Pi_\mu^\lambda \hat  \Pi_{\lambda \nu}+\frac12\eta_{\mu\nu}\left( \hat \Pi^2-\hat \Pi_{\mu}^\nu \hat \Pi_\nu^\mu\right)
 \Big]\,.
\eea
We can perform a field redefinition that decouples helicity 2 from helicity 0 fields 
\be
\hat{h}_{\mu\nu}\,\to\,\hat{h}_{\mu\nu}+(3-2 c_0) \eta_{\mu\nu} \hat{\pi}- \partial_{\mu} \hat{\pi} \partial_\nu \hat{\pi}\,.
\ee
Then the kinetic terms for tensor and scalar are diagonalized, resulting up to total derivatives and terms independent of fields
\bea
{\cal L}_{h_{\mu\nu},\;\pi}&=&-\frac12  \hat{h}_{\mu\nu} {\cal E}^{\mu\nu\alpha\beta}  \hat{h}_{\alpha\beta}
- 3\,\left[  \lambda_0 -   H^2-\left(1-c_0\right) \left(2-c_0\right)\right]\, \left( \hat{h}
+4 (3-2 c_0) \hat{\pi}+\hat{\pi} \Box \hat{\pi}\right)
\nonumber\\
&&-12    H^2 (3-2 c_0) \hat{ \pi}
+\left[\frac32 \left(3-2 c_0\right)^2-6  H^2\right]
\hat{\pi} \Box \hat{\pi}-\frac32 \left(3-2 c_0\right) \left(\partial \hat \pi \right)^2 \Box \hat \pi
 \nonumber\\
&&
-\frac12 \left(\partial \hat \pi \right)^2 \left[\left(\Box  \hat \pi \right)^2
- \left( \partial_\mu \partial_\nu \hat  \pi \partial^\mu \partial^\nu\hat \pi\right) \right]\label{lagdec3}
\,.\eea

Let us emphasize a crucial feature of the previous Lagrangian: the decoupling limit and diagonalization procedures provide a contribution to the kinetic term of the scalar fluctuations, which includes a term proportional to the curvature $H^2$ of space-time. This term is inherited from a third order interaction between tensor and scalar, which after shifting the metric like in eq.~(\ref{shiftquan}), is obtained by applying the relation (\ref{urel3}).
  Perturbing around curved space, the scalar component of the massive graviton acquires a kinetic term depending on
 space-time curvature. This was already noticed in
\cite{ArkaniHamed:2002sp}, in the context of the Fierz-Pauli theory.

 We then observe that the Lagrangian (\ref{lagdec3}) contains two tadpole terms that depend linearly on the fields. Therefore, in order to ensure that our configurations (\ref{metrdec1})-(\ref{stuckdec1}) are solutions of background field equations, we have to remove these tadpoles by setting their coefficients to zero. The conditions that we find are
\bea
 \lambda_0 -  H^2-\left(1-c_0\right) \left(2-c_0\right)&=&0  \label{ceq1}\,,\\
  H^2 (3-2 c_0) &=&0 \label{ceq2}\,.
\eea
The last of these conditions gives two branches of solutions. We will discuss these background solutions and their stability under cosmological perturbations in Section \ref{sec-mscsc} after analysing the dynamics of vector fluctuations in the next subsection.

\subsection{Including vectors fluctuations}\label{sec-msnawv}
 Here we discuss in detail the Lagrangian for vector fluctuations, the less explored sector of this theory.
We will follow and develop further the procedure originally elaborated in
  \cite{Koyama:2011wx}, that we briefly review now. In Section \ref{sec-prel}, we defined the tensor $H_{\mu\nu}$ as (recall that we set $\Lambda_3=1$)
\be
H_{\mu\nu}\,=\,g_{\mu\nu}-\left( \delta_\mu^\alpha-\Pi_\mu^\alpha-m\,\partial_\mu A^\alpha\right)\,\eta_{\alpha \beta}\,
\left( \delta_\nu^\beta-\Pi_\nu^\beta-m\,\partial_\nu A^\beta\right)\,.
\ee
From now on, we can neglect the helicity two fluctuation $h_{\mu\nu}$, since in the decoupling limit,  $h_{\mu\nu}$ does not couple to the vectors, hence all possible interactions and self-interactions are those studied in the the previous section. Recall that we raise and lower indexes with $\eta_{\mu\nu}$. It is convenient to adopt a matrix notation for tensors. If $B_{\mu\nu}$ and $C_{\mu\nu}$ are two tensors, we write
  $$B= B_\mu^{\;\;\nu} \;;\;  B C=  B_\mu^{\;\;\rho}  C_\rho^{\;\;\nu}  \;;\;  B C^T=  B_\mu^{\;\;\rho}  C_{\;\;\rho}^{\nu} \,  \;;\;\tr{B}\,=\,B_\mu^{\;\;\mu}\,\,.$$
Let us introduce the tensor $$M_{\mu}^{\;\;\nu}\,=\,\delta_\mu^{\;\;\nu}-H_\mu^{\;\;\nu}\,.$$
Then one can write
\bea
M\,=\,M_\mu^{\;\;\nu}
&=&\left(P^2\right)_{\mu}^{\;\;\nu}-m \left(L_{1}\right)_{ \,\mu}^{\;\;\nu}+m^2 \left(L_{2}\right)_{ \,\mu}^{\;\;\nu}\,=\,P^2-m L_1 +m^2 L_2
\eea
with ($d A\,\equiv\, \partial_\mu A^{\nu}$)
\bea
P_\mu^{\;\;\nu}&=&\delta_\mu^{\;\;\nu}-\Pi_\mu^{\;\;\nu}\,=\,{\bf 1}-\Pi\,,\nonumber \\
L_{1 \,\mu}^{\;\;\;\;\nu}&=&\partial_\mu A^\alpha \,P_\alpha^{\;\;\nu}+P_\mu^{\;\;\beta} \,\partial^\nu A_\beta\,=\, dA\, P+P \,dA^T\,,\nonumber \\
L_{2 \,\mu}^{\;\;\;\;\nu}&=&\partial_\mu A^\alpha \partial^\nu A_\alpha\,=\, dA\,dA^T\,.
\eea
We will need to take the square root of $M$ up to second order in an expansion in $m$, so it is convenient to write it as
\bea
M&=&P^2- m\,L_1+m^2\,L_2,
\\
&=&P^2\,\left[1-m\,Q_1+m^2\,Q_2 \right]^2
\eea
with
\be
Q_1=\frac12 P^{-2} L_1\,,\qquad
\qquad Q_2=\frac12 P^{-2} L_2-\frac18 P^{-2} L_1 P^{-2} L_1\,.
\ee

Then
\be
\sqrt{M}\,=\,P \,\left(1-m\,Q_1+m^2\,Q_2 \right)\,+\,m D+m^2 E +\mathcal{O}(m^3)
\ee
with $D$ and $E$ two matrices satisfying the following equations (see \cite{Koyama:2011wx} for details)
\bea\label{probeq}
 \left\{ D, \,P \right\}
&=&P  \left[ Q_1,\,P
\right],\\
\label{eqq2}
0&=&
\left[  Q_2 ,\,P\right]+\left[ D,\,Q_1\right]+ P^{-1}\,\left\{ P,\,E\right\} +P^{-1}\, D^2 ,
\eea
where $[,]$ and $\{,\}$ are the commutator and anticommutator respectively. Taking traces, one finds
\be
\tr{D}=0\,,\qquad \qquad \tr{E}=-\frac12\tr \left( P^{-1} D^2\right)\,.
\ee
The previous formulae are correct up to second order in an $m$ expansion, and this is sufficient for the decoupling limit, since higfher order terms vanish.

The 
vector-scalar Lagrangian that can be condensed in the following expression
\be
{\cal L}_{A_\mu}\,=\,-\frac{1}{m^2}\left[ 6\, \tr{\sqrt{M}}+\tr{M}-\left( \tr{\sqrt{M}}\right)^2-12\right]\,.
\ee
Plugging the previous formulae for $M$ and $\sqrt{M}$, and focussing on the part inside the parenthesis that is proportional to $m^2$ (term with $m^0$ and $m^1$ are total derivatives or field-independent, and those with higher powers in $m$ vanish in decoupling limit $m\to0$), one finds  \cite{Koyama:2011wx}
 \bea\label{lagvec1}
 {\cal L}_{A_\mu}&=&-\left\{
 \left[\tr L_2-( \tr P Q_1)^2\right]+2 \left(3-\, \tr P\right)\,\tr\left( P Q_2\right)+\left( \tr P -3\right)\,\tr \left( P^{-1} D^2\right)
 \right\}\,.
 \eea

We are now interested in the dynamics of fluctuations around the maximally symmetric configurations discussed in the previous section. In order to achieve this, we perform the shift (\ref{pishift1}) to the scalar background configuration, that amounts to express $P$ as
\be\label{shonp}
P =\left(c_0\,{\bf 1}-\hat{\Pi} \right)\,.
\ee
Because the above tensor $P$ is obtained by taking the square root of $P^2$, we are forced to set $c_0>0$.
We decide to calculate terms up to quartic in perturbations (in the hat quantities). The solution for the matrix equation (\ref{probeq}) determining $D$ starts quadratic in perturbations, and it reads at this order
  \be
  D\,=\,\frac{1}{c_0}\,\left[\hat{\Pi}, \left( d A+ d A^T\right) \right].
  \ee
 Calculating the remaining contributions and assembling them together, one finds, up to quartic order in perturbations, the following action coupling vectors to scalars

 \bea
{\cal L}_{A_\mu}&=&\frac{1}{4\,c_0} \Big\{ (3-2 c_0)\,\tr{F^2}
 +\tr \hat \Pi \,\tr F^2+\frac{(3-4 c_0)}{ c_0} \,\tr\left( \hat \Pi F^2\right)
 \\&&\frac{\left(3-4 c_0\right)}{2\, c_0^2} \left[\tr (\hat \Pi^2 F^2) +\tr (\hat \Pi F \hat \Pi F)\right]
  +\frac{1}{ c_0} \tr \hat \Pi \,\tr (\hat \Pi F^2)
  \Big\}
  \,+\,\dots\label{fvlag}
 \eea

This Lagrangian exhibits both a Galileon symmetry in the scalar sector, and an abelian gauge symmetry in the vector sector. The simple and symmetric form for this scalar-vector Lagrangian will allow us to study the stability of cosmological vacuum solutions in a straightforward manner. Notice that
 the previous terms are only the firsts of an infinite series that characterizes the vector Lagrangian in the decoupling
  limit \cite{Koyama:2011wx}. Our method allows one to systematically  compute each order in perturbations, in a covariant
  way that makes manifest the symmetries of the system.
   These initial terms are, in any case, sufficient for our purposes (see however Appendix \ref{app-resum} for a resummation of all the contributions in a special case).


\subsection{Maximally symmetric configurations and their stability}\label{sec-mscsc}
Armed with the previous results, let us now discuss the properties of the configurations that solve the two tadpole equations (\ref{ceq1}) and (\ref{ceq2}), and the the dynamics of fluctuations around these solutions. As we anticipated, eq.~(\ref{ceq2}) gives two branches of solutions (from now on, for simplicity, we will remove the hats from the scalar fluctuations):

\bigskip
\noindent
{\bf Branch I: Minkowski space.}
Choosing $H=0$ to solve  eq.~(\ref{ceq2})  we obtain   Minkowski space (the resulting configuration corresponds to the screening solution discussed in \cite{deRham:2010tw}). In this case,  the second tadpole eq.~(\ref{ceq1}) gives
\be\label{scc0}
c_0=\frac{3}{2}(1\mp \frac13 \sqrt{1+4 \lambda_0}),
\ee
which in order to have a well defined square root, forces $\lambda_0\ge -1/4$. Let us consider now the Lagrangian for vector fluctuations, eq.~(\ref{fvlag}), that we rewrite keeping terms up to third order in fluctuations and expanding the traces:
\bea
{\cal L}_{A_\mu}&=&-\frac{(3-2 c_0)}{4\,c_0}\,F_{\mu\nu}F^{\mu\nu}-\frac{1}{4\,c_0} \Big[
 \Box \pi \, F_{\mu\nu} F^{\mu\nu}\,+\,\frac{(3-4 c_0)}{ c_0} \, \partial_\mu \partial_\nu \pi F^{\mu \rho} F^{\nu}_{\;\;\rho }
  %
  \Big]\,.
  \label{fvlaga1}
 \eea
We notice that the kinetic term for the vector fluctuations is healthy when $c_0$ lies in the interval $0<c_0<3/2$. This implies that only the choice with minus sign in eq.~(\ref{scc0}) gives a healthy kinetic term for the vector,  and
 the cosmological constant $\lambda_0$ in this branch should be contained in the interval  $-1/4\le \lambda_0<2$ in order to avoid ghosts and have a well defined solution. The third order terms in the Lagrangian (\ref{fvlaga1}) can be removed by the following field redefinition of the vector field
\be
F_{\mu\nu}\,=\,\sqrt{\frac{c_0}{(3-2 c_0)}}\,\left[\hat{F}_{\mu\nu}+\frac{1}{2(2 c_0-3)}\, \Box  \pi \,  \hat{F}_{\mu\nu}+\frac{3-4 c_0}{2 \,c_0\,(2 c_0-3)}\,  \partial_\mu \partial_\rho \pi \,
\hat{F}_{\nu}^{\,\,\rho}\right]\label{fmnred}
\ee
This redefinition, valid as long as $c_0$ lies in the interval $0<c_0<3/2$, renders the vector Lagrangian canonical up to third order in perturbations.  Similar redefinitions should allow to remove  terms beyond the cubic order in the scalar-vector Lagrangian. 

\bigskip
\noindent
{\bf Branch II: (A)dS space.}
This amounts to choose $c_0=3/2$ to satisfy eq.~(\ref{ceq2}). Then, the second tadpole, eq.~(\ref{ceq1}), implies the condition $H^2 =\frac{1}{4}(1+4 \lambda_0)$.
By inspecting the kinetic term for the scalar in eq (\ref{lagdec3}), one then finds that, in order to avoid a
ghost in the scalar sector, the parameter $H^2$ must be negative (corresponding to AdS space). However, in this case $c_0=3/2$ the vector Lagrangian is strongly coupled,  since the vector kinetic term vanishes. The
%
  vector Lagrangian (\ref{fvlag}) reads in this case
\bea
{\cal L}_{A_\mu}&=&
\frac{1}{18} \left[3 \left( \tr \Pi\,\tr F^2 -2 \,\tr \Pi F^2 \right) - 2  \left(\tr \Pi^2 F^2  - \tr \Pi\, \tr \Pi F^2 + \tr \Pi F \Pi F\right)\right]
  \,+\,\dots\label{fvlags}
 \eea
 As mentioned above,
the kinetic terms for the vectors vanish; however, vectors become dynamical by coupling them with the scalar at third or higher order in fluctuations (this was already pointed out in \cite{deRham:2010tw,Koyama:2011wx,D'Amico:2012pi}). Nevertheless, one should worry about
higher derivatives in the equations of motion, since the previous Lagrangian contains contributions with two time derivatives in the scalar field $\pi$. For systems coupling scalars with vectors, it is possible to find the combination  that ensures that the equations of motion do not contain at all terms containing more than two time derivatives. It is a generalization of Galileon combinations which was explored in \cite{Deffayet:2010zh} and dubbed $p$-form Galileons. Up to fourth order in perturbations, the correct combination (without including higher derivatives in $A_\mu$)
  is
 \be
{\cal L}^{p-form}\,=\,a_0\,\left[ \tr \Pi\,\tr F^2 -2 \,\tr \Pi F^2 \right] +b_0\,\left\{
\tr F^2\,\left[\tr \Pi^2 -\left( \tr \Pi\right)^2\right]-4 \tr \Pi\, \tr \Pi F^2 + 4 \tr \Pi^2 F^2 +2  \tr \Pi F \Pi F
\right\}\label{pfgal}
\ee
where $a_0$ and $b_0$ are arbitrary coefficients. The above third order action with the aforementioned properties
was presented in \cite{Deffayet:2010zh}, while the fourth order one is as far as we know new.
 Comparing (\ref{fvlags}) with (\ref{pfgal}) we notice that
 while the third order action has the correct structure to avoid higher order time derivatives in the equation of motion,
 the fourth order Lagrangian does not seem to satisfy this requirement. However,  a suitable field redefinition allows
 to recast eq. (\ref{fvlags}) into a healthy form:
  %
   we will discuss  this technical point
   in
 Appendix  \ref{app-resum},
 also providing strong arguments that indicate that the same pattern continues at higher order in perturbations.

  %
 %

\bigskip

On the other hand, although our scalar-vector Lagrangian (\ref{fvlags}) does not lead to a propagation of a sixth ghost mode, it does
generally  lead
to a ghost-like instability around self-accelerating configurations, in which the ghost is one  of the available vector modes.
 We have already shown  in \cite{Koyama:2011wx} that, when turning on a non-trivial profile for the vector  solving the equations of motion,
 the corresponding Lagrangian for perturbations around the resulting configuration  acquires kinetic terms for the vector
 with the wrong sign. For completeness, we include a  proof of this fact in Appendix
\ref{app-vec}, based on the Lagrangian (\ref{fvlags})  expanded up to third order in perturbations.  In the remaining
 of this section, we instead directly point out the instability by  analyzing the hamiltonian associated with the Lagrangian
   obtained by combining the third order lagrangian contained in (\ref{fvlags}) with the scalar kinetic term:

\be\label{thirdoa2}
{\cal L}^{third}\,=\,-3  H^2
{\pi} \Box{\pi}-\frac16 \left[ \Box \pi\,F_{\mu\nu} F^{\mu\nu}-2 \partial_{\mu\nu} \pi F^{\mu\rho} F^{\nu}_{\;\;\rho}\right]
\ee
We choose, for simplicity, the gauge $A_0=0$, $\partial_i A_i=0$; by doing a standard $3+1-$decomposition, the previous Lagrangian reads
\be
{\cal L}\,=\,-3 H^2 \dot{\pi}^2+\frac13\,\left[2 \dot{\pi} \,\dot{A}_i  \triangle A_i+\triangle \pi \,\dot{A}_i^2-\pi_{,\,ij} \,\dot{A}_i \dot{A}_j\right]+\dots
\ee
where the dots represent the terms without time derivatives, which we do not include since they do not play a role in the present discussion.
The conjugate momenta to $\pi$ and $A_i$ are
\bea
\Pi_{\pi}&=&-6 H^2 \,\left(\dot{\pi}-\frac{1}{9\,H^2} \,\dot{A}_i  \triangle A_i\right)\label{conm1}\,,\\
\Pi_{A_i}&=&\frac{2}{3}\left[ \dot{\pi} \, \triangle A_i+\triangle \pi \,\dot{A}_i -\pi_{,\,ij} \,\dot{A}_j
\right]\label{conm2}\,.
\eea


In order to analyze the associated Hamiltonian, it is convenient to introduce the matrix
\be
{\kappa}_{ij}\,\equiv\,\triangle \pi \,\delta_{i j} -\pi_{,\,ij}
\ee
written in  terms of spatial derivatives of the scalar fluctuation; our analysis depends on the properties of this object. First, let us suppose
that the scalars fluctuations are such that this matrix vanishes: $\kappa_{ij}\,=\,0$. Then, we can easily invert the relations that define the
conjugate momenta, and obtain
\bea
\dot{\pi}&=&\frac{3\,A_i \Pi_{A_i}}{2\,A_i \,\triangle A_i}\\
\dot{A}_i \, \triangle A_i&=&\frac32\,\left(\Pi_\pi+\frac{9 H^2\,A_i \Pi_{A_i}}{\,A_i \,\triangle A_i} 
\right)
\eea
which translates into the following Hamiltonian
\bea
{\cal H}&=&-\frac{\Pi_\pi^2}{12 \,H^2}+\frac{1}{12 H^2}\, \left(
\Pi_\pi+\frac{9\,H^2\,A_i \Pi_{A_i}}{A_i \,\triangle A_i}
\right)^2+\dots\,,\\
&=&\frac{3}{2}\, \Pi_\pi\,\frac{A_i \Pi_{A_i}}{A_i \,\triangle A_i}+
\frac{27\,H^2}{4 }\, \left(
\frac{A_i \Pi_{A_i}}{A_i \,\triangle A_i}
\right)^2+\dots\,,
 \label{haminf2}
\eea
where the dots represent terms without momentum variables.
 The previous hamiltonian is linear in $\Pi_\pi$; hence it is unbounded from below. Notice that this argument holds
 even in the limit in which $H^2$ vanishes.
In conclusion, perturbations of the background
self-accelerating solution, along the direction of  scalar fluctuations such that ${\kappa}_{ij}\,=\,0$,
admit unstable directions along which
  the system falls towards regions where the energy is unbounded from below.

Similar conclusions hold for more generic $\kappa_{ij}$. Let us, for example, consider a $\kappa_{ij}$ that is non-vanishing, and invertible. Then, after straightforward manipulations, one can show that the Hamiltonian can be written as

\be
\frac43\,{\cal H}\,=\,-\frac{1}{9 H^2+\Delta A_i \kappa_{ij}^{-1} \Delta A_j}\,\left(\Pi_\pi-  \Delta A_i \kappa_{ij}^{-1}  \Pi_{A_j}\right)^2
+ \Pi_{A_i} \kappa_{ij}^{-1}  \Pi_{A_j}+\dots\,,
\ee
where, again, the dots represent terms without momentum variables.
It is not difficult to see that there are many unstable directions associated with this Hamiltonian.
 For example,  make a choice  for the vector $\Delta A_i$, so that the scalar combination ${\cal C}\,\equiv\,\Delta A_i \kappa_{ij}^{-1} \Delta A_j$ is non-vanishing and has a certain sign;
   for definiteness, the magnitude  of $\Delta A_i$ is chosen such that
the denominator of the first term has the same sign of ${\cal C}$.
 Accordingly, choose the vector $\Pi_{A_i}$ such that $ \Pi_{A_i} \kappa_{ij}^{-1} \Pi_{A_j} $ has the same sign of ${\cal C}$ (for example, choose it in the same direction of the $\Delta A_i$).  Then, choosing suitably the magnitude of $\Pi_\pi$,  it is possible to make one
of the two terms in the previous hamiltonian arbitrarily  negative -- hence the Hamiltonian is unbounded from below.
Other cases, such as the case in which $\kappa_{ij}^{-1}$ is non-vanishing but not invertible, can be treated in a similar way.

In conclusion, one generically expects that instabilities arise. There are many directions in the moduli space of fluctuations along
which the energy is unbounded from below, and towards which the
  system can be driven into
  dangerous regions.

\bigskip

\begin{figure}[htp!]
\begin{center}
\includegraphics[scale=1]{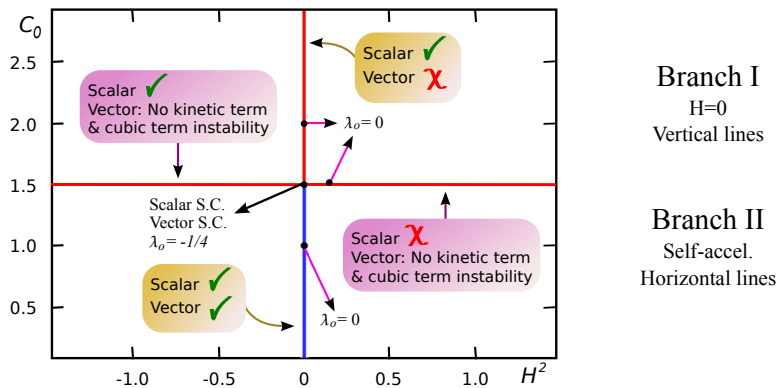}
\end{center}
\caption{  Branches I (solid vertical line with $H=0$) and II (solid horizontal line) solutions for $\alpha_3=\alpha_4=0$. Healty configurations are shown in blue. The branch II of self-accelerating solutions has no kinetic term for the vector field, and there are directions in field space where the Hamiltonian for perturbations is unbounded from below, due to the cubic interaction (\ref{thirdoa2}). Unstable solutions present ghosts either in the scalar sector (branch II) or in the vector sector (branch I and II), and are represented by red. In the branch intersection, there is strong coupling (S.C.) for both the vector and scalar modes, since they do not have kinetic terms,
 but the vector instability remains.
}
\label{fig1}
\end{figure}

To summarize, for each value of $\lambda_0$ it is possible to determine a unique configuration  that belongs to the first or second branch (see figure \ref{fig1}). Branch I are the screening solutions discussed in \cite{deRham:2010tw}, where the allowed range for the cosmological constant $\lambda_0$ is small, from $-1/4$ to $2$ in order to avoid ghosts.
The value $\lambda_0=-1/4$ is special, being common to both branches. In this case, scalar and vector degrees of freedom are strongly coupled.  This is the case studied in \cite{Tasinato:2012mf} in the context of Fierz-Pauli massive gravity, where new symmetries were observed in linear perturbations. It is not clear whether these symmetries survive to higher order in perturbation theory. However, from Lagrangian (\ref{fvlag}) one can see that in this common point, the cubic terms do not vanish, suggesting the disapperance of the symmetry beyond linear order in perturbations in  $\Lambda_3$ theory:
  in fact, as we have found above, these higher order contributions make the case $H=0$ unstable. For $\lambda_0<-1/4$, the cosmological constant curves the space leading to an AdS configuration that is unstable since it is characterized by a ghost in the vector
  or scalar sectors. As we anticipated, the analysis of vector fluctuations is essential for characterizing completely the two branches and exhibit the
instabilities.
In the next section, we repeat the analysis  using configurations with non-vanishing $\alpha_3$ and $\alpha_4$.


\section{Maximal symmetric solutions and  instabilities  for  arbitrary $\alpha_3$ and $\alpha_4$}\label{sec-mswa}
In this section, we extend the previous discussion to the case of arbitrary $\alpha_3$ and $\alpha_4$.  Including the contributions corresponding to non-vanishing $\alpha_3$ and $\alpha_4$, the Lagrangian in the decoupling limit reads
\bea
{\cal L}_{dec}&=&-\frac12 \,h^{\mu \nu} {\cal E}_{\mu \nu}^{\alpha \beta} h_{\alpha\beta}
- 3 \,\lambda_0\, h +h^{\mu\nu} \left( {X}^{(1)}_{\mu\nu} +{X}^{(2)}_{\mu\nu} +{X}^{(3)}_{\mu\nu} \right)
\eea
The expression for the $X^{(i)}_{\mu\nu}$ is given in eqs (\ref{xiexp}).
 To study this Lagrangian and its maximally symmetric solutions, we proceed exactly as in section \ref{sec-msna}, performing the
shifts of tensor and scalar modes as in eqs (\ref{shiftquan})-(\ref{pishift1}). We  skip the
details as they are a straightforward generalization of the arguments of the previous section, and present only the final results.

In order to eliminate the terms that are linear in the fields in the resulting Lagrangian (tadpoles), one obtains the following set of two equations that generalize eqs (\ref{ceq1})-(\ref{ceq2}):
\bea
  \lambda_0 -   H^2-\left(1-c_0\right) \left[2 -c_0+ 4 \alpha_4 (1 - c_0)^2
  + \alpha_3 \left(4 - 5 c_0 + c_0^2\right)\right]&=&0  \label{ceq1b}\,,\\
  H^2 \left[3 - 2 c_0 + 12 \alpha_4 (1 - c_0)^2 + 3 \alpha_3 (3 - 4 c_0 + c_0^2)\right] &=&0 \label{ceq2b}\,.
\eea
After diagonalizing the Lagrangian (although at fourth order in perturbations a mixing between $h$ and $\pi$ remains, when
$\alpha_3\neq-4\alpha_4$), one obtains the following
  kinetic term for the scalar field
\bea
{\cal L}_{kin,\; \pi}&=&
\Big\{\frac32\,\left[3 - 2 c_0 + 12 \alpha_4 (1 - c_0)^2 + 3 \alpha_3 (3 - 4 c_0 + c_0^2)\right]^2\nonumber\\
&&\hskip1cm-3 H^2\,\left[1 + 3 \alpha_3 (2 - c_0) + 12 \alpha_4 (1 - c_0) \right]  \Big\}
\pi \Box \pi\,.
\eea
 These results were already obtained in \cite{deRham:2010tw}. The new piece of information is the Lagrangian for vector perturbations, which up to third order in perturbations becomes (obtained following the same method of section \ref{sec-msnawv})
 \bea
 {\cal L}_{A_\mu}&=&\frac{\left[3+12{\alpha_4} ({c_0}-1)^2-2 {c_0}+3 {\alpha_3} (3+({c_0}-4) {c_0})\right] {\tr{F}_2}}{4 {c_0}}\nonumber\\
 &&
- \frac{\left[-1+3 \alpha_3 (-2+c_0) +12 \alpha_4 (-1 + c_0) \right]  \,{\tr\Pi}\,\tr{F}_2 }{4 {c_0}} \label{lvcomp}
\\
 &&+\frac{1}{4 {c_0}^2}\left[
 3 + 9 \alpha_3+12 \alpha_4 - 4 (1+6 \alpha_3+12 \alpha_4) c_0+9 (\alpha_3+4 \alpha_4) c_0^2 \right] \tr{\Pi\,F_2}\,.
  \nonumber
 \eea
The previous Lagrangian contains only the firsts  of an infinite series of terms including vectors, but the terms above are enough for our purposes.

\bigskip
Within this system, we would like to determine maximally symmetric configurations and discuss their stability. As in the previous section, eq.~(\ref{ceq2b}) admits two branches of solutions: one corresponding to Minkowski space, and the other to a maximally symmetric space-time.
These are the generalization of the two branches of solutions discussed in section \ref{sec-mscsc}. Since we are interested in characterising self-accelerating de Sitter configurations, we will not discuss the Minkowski branch any further in the main text, relegating its discussion to Appendix \ref{app-fb}. We will, instead, focus on the maximally symmetric option $H^2\neq 0$ for solving eq.~(\ref{ceq2b}). The solution
 for $c_0$ is then
 \bea
 c_0^\pm&=&\frac{\left(1+3 \a_3\right)+3 \left(\a_3+4 \a_4\right)\pm\sqrt{ \left(1+3 \a_3\right)^2-3 \left(\a_3+4 \a_4\right)}}{3\left( \a_3+4 \a_4\right)}\label{rc01}\,,
 \\ \label{rc02}
 &=&
 \frac{2+3\alpha_3\mp | \alpha_5|}{1+3\alpha_3\mp | \alpha_5|}\label{c0opt}\,.
 \eea
where, in order to render the formulae less cumbersome, we define (as done in \cite{Koyama:2011wx})
  \bea
  \alpha_5^2& \equiv&\left(1+3\a_3\right)^2-3 \left(\a_3+4 \a_4\right)\,.
  \eea
Notice that in the special case of $\alpha_3=-4\alpha_4$, the positive branch $c_0^+$ disappears and only $c_0^-$ remains physical. 
In the following, we will impose $\alpha_5^2\ge 0$ in order to have well-defined square root in eq.~(\ref{rc01}), and $c_0>0$ as discussed after eq.~(\ref{shonp}).
Upon substituting the previous values for $c_0$ in the expression for the scalar kinetic terms, we find
 \be\label{kinpige}
 {\cal L}_{kin,\pi}\,=\,\pm\, 3\, |\alpha_5|\, H^2\,\pi \Box \pi\,.
 \ee
Substituting the value of $c_0$ from eq.~(\ref{rc02}) into eq.~(\ref{ceq1b}), we find the following expression for $H^2$
\be
H^2\,=\, H^\pm_0+  \lambda_0 \,,\qquad\qquad  H^\pm_0 \equiv\,
%
\frac{1}{3}\frac{1+3\alpha_3\mp 2| \alpha_5|}{(1+3\alpha_3\mp | \alpha_5|)^2}\, .\label{solh2}
\ee
The  results so far suggests that self-accelerating configurations with $H^2>0$ may be allowed in this theory, by a suitably choice of $\alpha_3$ and $\alpha_5$; this is indeed the result of \cite{deRham:2010tw}.  However, let us  see what happens to vector fluctuations.
The coefficient in front of the vector kinetic term can be written in terms of $c_0^{\pm}$ as
\be
{\cal L}_{kin,A_{\mu}}
=  \frac{ \left[(1+3\alpha_3)^2-\alpha_5^2\right]}{4 c_0} (c_0 - c_0^+)(c_0 - c_0^-) \tr F_2.
\ee
The solutions for $c_0$, eq.~(\ref{rc02}), automatically imply that the vector fluctuations are strongly coupled. After substituting our solution for $c_0$ of eq.~(\ref{rc02}) in the Lagrangian (\ref{lvcomp}), we find
\bea
  {\cal L}_{A_\mu, c_0=c_0^{\pm}}
 &=&\nonumber
 \mp |\alpha_5|\frac{(1+3 \alpha_3 \mp |\alpha_5|)}{4(2+3 \alpha_3 \mp |\alpha_5|)}
  \,\Big[ {\tr\Pi}\,\tr{F}_2 -2\, \tr{\left( \Pi\,F_2\right)} \Big]\\
 &=&\mp \frac{ |\alpha_5|}{4c_0^{\pm}}
  \,\Big[ {\tr\Pi}\,\tr{F}_2 -2\, \tr{\left( \Pi\,F_2\right)} \Big]\label{lvcomp22}
 \eea
This is a generalisation of the result for $\alpha_3=\alpha_4=0$.
This Lagrangian leads to equations of motion
that are at most second order in time derivatives.
 However,
using  arguments  identical
 to the ones developed in section \ref{sec-mscsc},  it is possible to see that, when $\alpha_5\neq 0$,  the Hamiltonian associated with this system is unbounded from below if one moves along certain directions in the moduli space of perturbations. Hence,
 for the very same reasons explained in section  \ref{sec-mscsc},  also these solutions with arbitrary $\alpha_3$ and $\alpha_4$
  (and keeping $\alpha_5\neq0$) are unstable
 when $c_0$ satisfies eq. (\ref{rc02}).
In Fig. 2 we summarise the stability of the solutions in the general case. The special case $\alpha_5=0$ is more subtle, since
both scalar and third order vector Lagrangian vanish, and it is related with the configurations
analyzed in \cite{sbh}. We will postpone its study to the future.

\begin{figure}[htp!]
\begin{center}
\includegraphics[scale=1]{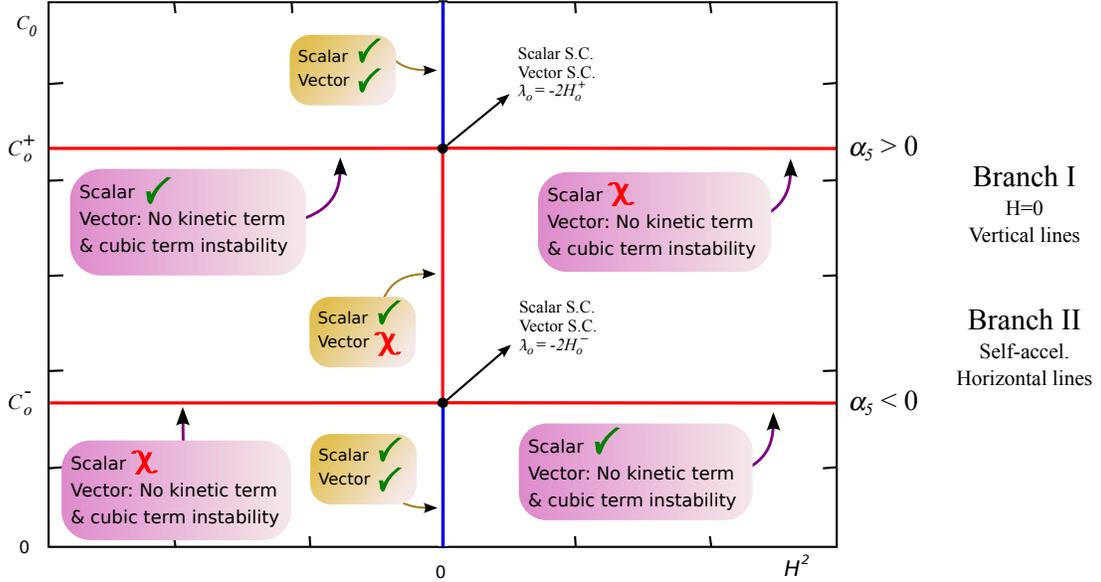} \hskip .5cm
\end{center}
\caption{
 Stability analysis of screening solutions (branch I) and self-accelerating solutions (branch II) for a general $\alpha_3$ and $\alpha_4$, obeying the condition $(1+3\alpha_3)^2 > \alpha_5^2$. Solutions with instabilities are shown in red and healthy ones in blue. The precise values of $H_0^\pm$, $c_0^\pm$ and the points where $\lambda_0=0$ depend on the explicit values of $\alpha_3$ and $\alpha_4$. There are two branches of self-accelerating solutions (solid horizontal lines), which lack of kinetic terms for vector perturbations (strong coupling), and present an instability when cubic interactions are considered. If $\alpha_3$ and $\alpha_5$ are chosen correctly, one may have expected a healthy model of de-Sitter without a cosmological constant $\lambda_0$, but this is not the case due to the instabilities described in the main text. There are three branches of screening solutions with $H^2=0$ (solid vertical lines), which are connected at $c_0 =c_0^{\pm}$. In the case of $(1+3\alpha_3)^2<\alpha_5^2$, the
diagram is similar, but the screening solutions are healthy in the intermediate segment (between $c_0^+$ and $c_0^-$) and unstable elsewhere.
Finally, if $(1+3\alpha_3)^2=\alpha_5^2$, or equivalently $\alpha_3=-4\alpha_4$, the two self-accelerating branches collapse into a single one and it becomes similar to Fig.~\ref{fig1}.}
\label{fig2}
\end{figure}

\section{Discussion}\label{sec-disc}

In this work we studied, in detail, the dynamics of vector fluctuations in the decoupling limit of $\Lambda_3$ massive gravity, with the main aim of investigating the stability of self-accelerating configurations.  The decoupling limit of
  $\Lambda_3$ massive gravity propagates 5
physical degrees of freedom around any background: 2 tensors, 2 vectors, and 1 scalar.
While it is well known that, in the decoupling limit,  scalar fluctuations organize so to form the few scalar Galileon combinations, the structure of the vector Lagrangian is more subtle. It contains an infinite number of terms, one at each order in perturbations. We  developed a method that allows to determine, in a fully covariant way, the vector Lagrangian at each order in perturbations.
 Around self-accelerating configurations, the vectors become strongly coupled, being
 characterized by vanishing kinetic terms. This has been shown perturbatively in \cite{deRham:2010tw} and non-perturbatively for spherically symmetric perturbations in \cite{Koyama:2011wx}. In this paper, we showed non-perturbatively that the strong coupling of the vector perturbations is an
 inevitable consequence of selecting the self-accelerating branch of the solutions.
 However, the vector acquire dynamics thanks to contributions at third  and higher order in perturbations;
 at third order they form a $p$-form Galileon combination, which ensures, at this order, that the associated equations of motion contain
 at most two time derivatives. At fourth and higher orders it is not automatically so, and the equations of motion contain higher derivatives. However these higher derivatives are harmless, since one can show that field redefinitions recast the Lagrangian
 in a manifestly healthy form, which does not propagate additional degrees of freedom.
  We find interesting to reveal this structure of the decoupling limit Lagrangian when vectors are included. It generalises the standard scalar Galileon combinations to $p$-form Galileons and indicates in, yet another way, how the BD 'sixth mode' is not present
   around special backgrounds.

However, the strong coupling of the vector degrees of freedom around self-accelerating solutions is problematic, since it may turn one of the available five degrees of freedom into a ghost, making these configurations generically unstable. A careful Hamiltonian analysis of the system, indeed, reveals that the Hamiltonian is unbounded from below along generic directions in the space of fluctuations. It would be very interesting to understand whether such instability can be avoided, for example imposing symmetries that prevent to move along dangerous directions in moduli space.

The strong coupling also means that quantum corrections are not controllable in general, and we should stress that our analysis here is purely classical. It is known that the Galileon terms, without a vector, are protected from quantum corrections \cite{Hinterbichler:2010xn}, though we expect that the Maxwell term for vector perturbations will receive quantum corrections \cite{deRham:2010tw}. It will be important to study quantum corrections when vectors are included.

In our work we did not only limit our attention to self-accelerating solutions, but also to more general maximally symmetric solutions   in the presence of a bare cosmological constant. We investigated the two branches of vacuum solutions. The first one contains the self-accelerating configurations; it is characterized by a non-vanishing curvature $H^2$, the strong coupling of the vector sector, and aforementioned instability. The second branch is such that curvature $H^2$ vanishes, even in the presence of a bare cosmological constant. These solutions realise the screening configurations already studied in \cite{deRham:2010tw}. We confirm that, around these solutions, all fluctuations (vector included) are well behaved, provided the bare cosmological constant lies within an interval.

\smallskip

Our conclusions have been obtained in the decoupling limit, that corresponds to an analysis of the theory in a well defined, but limited, regime of scales. It would be interesting to reproduce and interpret these results from the full theory perspective, and point out other instabilities besides the ones analyzed in \cite{DeFelice:2012mx}. Such analysis will be carried on in a further publication \cite{new}, aimed also to clarify issues associated with the dynamics of scalar fluctuations in the full theory, which were found to exhibit a strong coupling behavior (no kinetic terms) when expanded at quadratic order around certain cosmological backgrounds \cite{Gumrukcuoglu:2011zh,D'Amico:2012pi}.

\acknowledgments

We would like to thank Guido D'Amico,  Emir G\"umr\"uk\c{c}\"uo\u{g}lu, Wayne Hu,  Nima Khosravi, Shinji Mukohyama and  Fulvio  Sbis\`a
for useful discussions. GT is supported by an STFC Advanced Fellowship ST/H005498/1.
KK is supported by STFC grant ST/H002774/1, the ERC and the Leverhulme trust.

 \begin{appendix}
\section{Higher derivatives and the scalar-vector Lagrangian}\label{app-resum}

In this appendix we investigate  the strongly coupled
 scalar vector Lagrangian of eq. (\ref{fvlags}), that we rewrite here:
 \bea
 {\cal L}_{A_\mu}&=&  {\cal L}_{A_\mu}^{(3)}+{\cal L}_{A_\mu}^{(4)}
 \nonumber\\&=&
\frac{1}{6}  \left( \tr \Pi\,\tr F^2 -2 \,\tr \Pi F^2 \right)
 - \frac19  \left(\tr \Pi^2 F^2  - \tr \Pi\, \tr \Pi F^2 + \tr \Pi F \Pi F\right)
  \,+\,\dots\label{fvlags5},
 \eea
 where the dots represent higher order perturbations.
  The  vector has no kinetic terms, and its dynamics is controlled by perturbations starting at third order (as
 already noticed in \cite{deRham:2010tw,Koyama:2011wx,D'Amico:2012pi}). The third order Lagrangian $ {\cal L}_{A_\mu}^{(3)}$
has a $p$-form Galileon structure \cite{Deffayet:2010zh}, and leads to equations of motion that contain at most two time derivatives:
 \bea
 \left(F_{\mu \nu,\,\rho}\right)^2-2 \left(F_{\mu \nu}^{\;\;\;\;,\nu}\right)^2\,=\,0\,,\\
 F^{\lambda \mu,\,\nu}\, \Pi_{\mu\nu}+F^{\mu \nu}_{\;\;\;\;,\nu}\,\Pi^{\lambda}_{\;\mu}-F^{\lambda \mu}_{\;\;\;\;,\mu}
 \Pi\,=\,0\,.
 \eea
 More delicate and interesting is the dynamics associated with the fourth order Lagrangian $ {\cal L}_{A_\mu}^{(4)}$. This has {\it
 not} the structure of a $p$-form Galileon (see eq.~(\ref{pfgal})), that in our context would read
 \be
 {\cal L}^{p-form}\,=\,\frac16\,\left(\tr \Pi\,\tr F^2 -2 \,\tr \Pi F^2 \right) +b_0\,\left\{\frac12
\tr F^2\,\left[\tr \Pi^2 -\left( \tr \Pi\right)^2\right]-2 \,\tr \Pi\, \tr \Pi F^2 + 2 \,\tr \Pi^2 F^2 +  \tr \Pi F \Pi F
\right\},\label{pfgal2}
 \ee
 with $b_0$ an arbitrary coefficient;
 then
   one might wonder whether it leads to additional degrees of freedom.
  We have many independent proofs, obtained following different procedures \cite{ghost} that show that
   $\Lambda_3$ massive gravity does not propagate a sixth ghost mode around any background. In the present context, we can
   make this fact concretely manifest in various ways.  For example, by showing that combining the equations of motion, it is possible to recast them
   in such a way that no higher derivatives are present. Or, alternatively,
    determining a field redefinition that, starting from the  Lagrangian of eq. (\ref{fvlags5}), allow to recast the 4th order part in
    a  form that coincides with the $p$-form Galileon of eq. (\ref{pfgal2}),
    so to render the system manifestly healthy. Let us  discuss  this second approach;  the field redefinition that we need is
   \be\label{nlfr}
   F\,\to\,F- \frac13\,\Pi\, F. 
   \ee
   By plugging the previous redefinition into the ${\cal L}_3$ part of (\ref{fvlags5}), we
   completely remove the fourth order part of the Lagragian (in other words, we
    generate  the healthy
   scalar-vector
   lagrangian as in eq. (\ref{pfgal2}) with $b_0=0$).

   On the other hand,   we know that our vector-scalar lagrangian contains an infinite number of contributions, while until now we only
   considered perturbations up to fourth order. What happens at higher order can be investigated by focussing on a simple, but
   representative example,
  %
      in which
    the total lagrangian can be computed at all orders and resummed.
%
  %
 %
%
%
We consider a situation in which only the following components of the vector $A_\mu$ are switched on:
 \be
 A_{\mu}\,=\,\left(A_0 (y),\,0,\,A_y (t),\,0\right)\,.
 \ee
We also suppose, again for simplicity,
  that the scalar depends only on the $(t,\,x,\,y)$ coordinates, and the cross derivatives vanish ($\Pi_{\mu\nu}=0$ if $\mu\neq\nu$). Then the action for the vector can be calculated {\it at all orders
  in perturbations} and can be resummed exactly  (as done
   in \cite{Koyama:2011wx}) finding for $c_0=3/2$
 \be
{\cal L}_{A_\mu}\,=\,\frac{F_{t y}^2\,\, {\Pi}_{xx}}{3+{\Pi}_{tt}-{\Pi}_{yy}}\,.
 \ee
 where $F_{yt}\,=\,\left( \partial_y A_0-\partial_t A_y\right)$.
 Notice that second derivatives appear at the denominator; such a system would naively seem to lead to the propagation of additional modes. This Lagrangian  has the same  structure of the one discussed
 in \cite{Koyama:2011wx} for the spherically symmetric case, and more recently in \cite{Gabadadze:2012he}
 for a purely scalar theory. Since we have only one component $F_{t y}$ switched on for the tensor $F_{\mu\nu}$
 it is simple in this case to exhibit a field redefinition (that generalizes (\ref{nlfr}) at higher orders) that removes the dangerous terms from
 the Lagrangian. More instructive is in this case to analyze the equations of motion, and showing that
 %
 %
%
 %
%
 by  manipulating them  they can recast in a form that manifestly does not lead to the propagation of an additional mode.
%
   The equation of motion for $A_y$ tells us that 
 \be
 \partial_{t}\left( \frac{\left( \partial_y A_0-\partial_t A_y\right) {\Pi}_{xx}}{3+{\Pi}_{tt}-{\Pi}_{yy}}\right)\,=\,0
 \hskip1cm \Rightarrow \hskip1cm\left( \frac{\left( \partial_y A_0-\partial_t A_y\right) {\Pi}_{xx}}{3+{\Pi}_{tt}-{\Pi}_{yy}}\right)\,=\,\text{constant}\label{const}\,\,.
 \ee
 The equation  of motion of $\pi$
 gives
 \be\label{prob1}
 \partial_t^2\left( \frac{\left( \partial_y A_0-\partial_t A_y\right)^2 {\Pi}_{xx}}{\left(3+{\Pi}_{tt}-{\Pi}_{yy}\right)^2}\right)
 \,+\,\text{parts that contain at most two time derivs}\,=\,0\,\,.
 \ee
Plugging (\ref{const}) into (\ref{prob1}), one easily  finds that all terms containing three or more time derivatives cancel out. Consequently, these higher time derivatives terms do not lead to additional 'ghost' mode.
 This shows directly, at least within this special ansatz, that the resummed Lagrangian does not propagate additional degrees of freedom.
  It would be very interesting to extend these results  to the complete Lagrangian for
  the vector-scalar modes,  without making any assumption on the field profiles.

\section{Turning on a background profile for the vector field}\label{app-vec}

In this section, we turn on a background profile for the vector field, and show that perturbations around the resulting configuration
always contain a ghost degree of freedom. The discussion essentially repeats the analysis of \cite{Koyama:2011wx}, although in the present set-up we make use of the Lagrangian only up to third order in perturbations (while in \cite{Koyama:2011wx} we considered a resummed Lagrangian in a spherically symmetric set-up).
We then focus on the following scalar-vector Lagrangian up to third order in perturbations
\be\label{thirdoa}
{\cal L}^{third}\,=\,-3  H^2
{\pi} \Box{\pi}-\frac16 \left[ \Box \pi\,F_{\mu\nu} F^{\mu\nu}-2 \partial_{\mu} \partial_{\nu}  \pi F^{\mu\rho} F^{\nu}_{\;\;\rho}\right]\,\,.
\ee
It is not hard to check that the corresponding equations of motion admit the following solution
\bea
{\pi}^{sol}&=&\frac{2\,Q_0^2}{3 H^2}\,t^2\,,\\
A_0^{sol}&=&-Q_0 \left(x^2+y^2+z^2\right)\,,
\eea
this configuration does not break the de Sitter symmetry of the physical metric.
 Perturbing the {third} order action around this solution, and calling the fields
\bea
\hat{\pi}&=&\frac{2\,Q_0^2}{3 H^2}\,t^2+\tilde{\pi}\,,\\
A_0&=&-Q_0 \left(x^2+y^2+z^2\right)+\tilde{A}_0\,,
\eea
we obtain (using the divergenceless condition $\partial_t A_0\,=\,\partial_i A_i$)  the following induced kinetic terms
at second order in the small perturbations $\tilde\pi$, $\tilde A_0$ \bea
{\cal L}^{\text{2nd order}}\,=\,-3  H^2
\tilde{\pi} \Box \tilde{\pi}+\frac{8\,Q_0}{3} \tilde\pi \Box \tilde A_0+\frac{4\,Q_0^2}{9 H^2} \tilde{F}_{ij}^2
\eea
The last term in the previous expression does not contain time derivatives.
Let us now focus on the time derivatives contained in  the first two terms, which read
\bea
{\cal L}^{\text{2nd order}}&=& 3  H^2
\tilde{\pi} \partial_t^2 \tilde{\pi}-\frac{8\,Q_0}{3} \tilde\pi \partial_t^2 \tilde A_0 +\text{pieces with no time derivatives}
\\
&=&3 H^2\,\left( \tilde \pi-\frac{4\,Q_0}{9 H^2} \tilde A_0\right)\,\partial_t^2\,\left( \tilde \pi-\frac{4\,Q_0}{9 H^2} \tilde A_0\right)-
\frac{16\, Q_0^2}{27\, H^2}\, \tilde A_0\, \partial_t^2\, A_0\,.
\eea
Defining an effective scalar $\bar \pi\,=\,\tilde \pi-\frac{4\,Q_0}{9 H^2} \tilde A_0$ one then finds that the previous Lagrangian contains always
a ghost when $H \neq 0$
 in agreement with \cite{Koyama:2011wx}. While in this appendix we focussed on the case $\alpha_3=\alpha_4=0$
it is straightforward to extend this analysis to the more general case, obtaining the same conclusion.

 \section{Minkowski branch of the solutions for arbitrary $\alpha_3$ and $\alpha_4$ }\label{app-fb}
In this appendix, we  discuss the solutions to eqs (\ref{ceq1b})-(\ref{ceq2b}) when selecting $H^2=0$ to solve eq.~(\ref{ceq2b}). In this case, both scalar and vector perturbation have non-vanishing kinetic terms, so it is sufficient to study the sign of their coefficients to investigate the stability of the configurations.
Using $c_0^{\pm}$ defined in (\ref{rc02}), the kinetic terms can be written as
\be
{\cal L}_{kin, \pi}
= \frac{27}{2} (\alpha_3 + 4 \alpha_4)^2 \Big[
(c_0- c_0^{+}) (c_0 - c_0^-)
\Big]^2 \pi \Box \pi,
\ee
\be
{\cal L}_{kin,A_{\mu}}
= \frac{3 (\alpha_3 + 4 \alpha_4)}{4 c_0} (c_0 - c_0^+)(c_0 - c_0^-)\, \tr F_2,
\ee
where $c_0$ is a solution of the tadpole equation (\ref{ceq1b}), which can be written as
\be \label{tadapp}
F(c_0) \equiv 2 + 4 \alpha_3 + 4 \alpha_4 -
 3 (1 + 3 \alpha_3 + 4 \alpha_4) c_0 + (1 + 6 \alpha_3 + 12 \alpha_4) c_0^2 - (\alpha_3+ 4\alpha_4) c_0^3=\lambda_0.
\ee
The stability of vector perturbations depend on the solution for $c_0$, which is a cubic equation in $c_0$ if $\alpha_3 \neq -4 \alpha_4$. The cubic function $F(c_0)$ has two stationary points at $c_0=c_0^{\pm}$, $F'(c_0^{\pm})=0$ where $\lambda_0 = \lambda_0^{\pm} = - 2 H_0^{\pm}$ ($H_0^{\pm}$ is defined in (\ref{solh2})). Thus there are three branches of solutions and at the stationary points $c_0=c_0^{\pm}$, where the two branches of solutions meet. These points are also the boundaries of the different stability behaviours. For example, if $\alpha_3>-4\alpha_4$, and $\alpha_3$ and $\alpha_4$ are such that $C_0^\pm>0$, then the vector is healthy between $c_0>c_0^+$ and $0<c_0<c_0^-$ (see figure \ref{fig2}). On the contrary, if $\alpha_3<-4\alpha_4$, but $c_0^\pm>0$, then the vector is healthy only within the region $c_0^+<c_0<c_0^-$.
For $\alpha_3=-4\alpha_4$, eq.~(\ref{tadapp}) is satisfied when
 \be\label{solgc0}
c_0\,=\,\frac{ 3+ 6 \alpha_3 \pm \sqrt{1 + 4 \,\left(1+3\alpha_3\right)\, \lambda_0}}{2 \left(1+3\alpha_3\right) }\,.
\ee
The kinetic terms for scalar and vector seem to vanish when  $\alpha_3=-4\alpha_4$; however it is not really the case since
$c_0^{\pm}$ have $(\alpha_3+4 \alpha_4)$ at denominator (see eq.~(\ref{rc01})).  Depending on the sign chosen in eq.~(\ref{solgc0}), the kinetic term for the vector can then be rewritten as
\be\label{comkinamu}
{\cal L}_{kin, A_\mu}\,=\,
\pm\frac{\left(1+3\alpha_3\right)\,  \sqrt{1+4 \,\left(1+3\alpha_3\right)\,  \lambda_0}}{2(3+6 \alpha_3\pm  \sqrt{1+4\, \left(1+3\alpha_3\right) \lambda_0})}\, F_{\mu\nu} F^{\mu\nu}
\ee
while the kinetic term for the scalar has no risk to become negative definite. Focussing for simplicity on the case $1+3\alpha_3>0$, the plus sign in eq (\ref{solgc0}) leads to a ghost in the vector sector. So only the negative sign is allowed, and the allowed solution for $c_0$ is
\be\label{solgc1}
c_0\,=\,\frac{ 1 + 2 \left(1+3\alpha_3\right) - \sqrt{1 + 4\,\left(1+3\alpha_3\right)\, \lambda_0}}{2 \left(1+3\alpha_3\right)\,}\,.
\ee
With this choice of sign, we have a well defined square root in (\ref{solgc1}) and a healthy kinetic term in (\ref{comkinamu}) if $\lambda_0$ lies within the interval
\be
-\frac{1}{4\,\left(1+3\alpha_3\right)\,}\,\le\,\lambda_0\,\le\,2+3\alpha_3\,\,,
\ee
generalizing the result discussed in Section \ref{sec-mscsc} obtained in the case $\alpha_3\,=\,\alpha_4\,=\,0$.





 \end{appendix}

\end{document}